\documentclass[prl,twocolumn,aps,amsmath,amssymb,superscriptaddress,10pt]{revtex4-1}
\usepackage{amsmath}
\usepackage{graphicx}

\newcommand{\bsym}[1]{\boldsymbol{#1}}
\newcommand{\ud}{\mathrm{d}}

\newcommand{\af}{JASLab, University of Ottawa and National Research Council, 100 Sussex Drive, Ottawa, Canada}
\newcommand{\aff}{Department of Electrical and Computer Engineering, University of New Mexico, Albuquerque, USA}
\newcommand{\afff}{Department of Physics, Texas A \& M University, College Station, USA}

\begin{document}

\title{Alignment dependent enhancement of the photo-electron cutoff for multiphoton ionization of molecules}

\author{C T L Smeenk}
\altaffiliation[Current address:]{Max Born Institute, Max-Born-Str. 2A, 12489 Berlin, Germany}
\affiliation{\af}
\author{L Arissian}
\affiliation{\aff}
\author{A V Sokolov}
\affiliation{\afff}
\author{M Spanner}
\affiliation{\af}
\author{K F Lee}
\affiliation{\af}
\affiliation{\afff}
\author{A Staudte}
\affiliation{\af}
\author{D M Villeneuve}
\affiliation{\af}
\author{P B Corkum}
\affiliation{\af}

\begin{abstract}

The multiphoton ionization rate of molecules depends on the alignment of the molecular axis with respect to the ionizing laser polarization.  By studying molecular frame photo-electron angular distributions from N$_2$, O$_2$ and benzene, we illustrate how the angle-dependent ionization rate affects the photo-electron cutoff energy. We find alignment can enhance the high energy cutoff of the photo-electron spectrum when probing along a nodal plane or when ionization is otherwise suppressed. This is supported by calculations using a tunneling model with a single ion state.
\end{abstract}
\maketitle

Ionization of atoms, molecules and solids by an intense, infrared laser pulse is frequently described as a tunneling process. In the gas phase, the tunneling model works well for describing the ionization rate  and photo-electron spectrum  of rare gas atoms \cite{Walker1994,Arissian2010}. Laser tunneling of molecules is generally more complex. In unaligned samples, the symmetry of the highest occupied molecular orbital can suppress the total ionization rate \cite{MuthBohm2000,Tong2002}. When the sample is aligned, the angle dependent tunneling current and location of nodal planes are imprinted onto the photo-electron spectrum \cite{Meckel2008,Akagi2009,Holmegaard2010}. The emitted electron wavefunction is then driven by the ionizing laser field, forming the basis for numerous attosecond spectroscopies that probe nuclear and electron dynamics \cite{Baker2006,Smirnova2009a}.  

The spatial resolution of the probing electron depends on its de Broglie wavelength $\lambda_{dB}=2 \pi/p$ (in atomic units). In laser tunneling, the electron momentum $p=E(t_i)/\omega$ where $E(t_i)$ is the laser field at the time of ionization and $\omega$ is the laser angular frequency  \cite{Corkum1989}. Therefore, at a constant frequency, the maximum photo-electron momentum is limited by the electric field an ensemble of atoms or molecules can survive before they are completely ionized. This is the saturation intensity. The saturation limit to the laser electric field in turn places a limit on the photo-electron's de Broglie wavelength in re-scattering processes like laser induced electron diffraction and high-harmonic generation \cite{Blaga2012,Shan2002}. These limits can be modified by changing the laser frequency or pulse duration \cite{Schmidt2011}.

Our approach is to control the ionizing medium rather than the light field. By using a pump laser pulse to align a molecular gas, we demonstrate an alternative approach to enhancement of the electron high energy cutoff. We study photo-electron momentum spectra from tunnel ionization of N$_2$, O$_2$ and C$_6$H$_6$ in circularly polarized, 800 nm laser pulses.  Circular polarization allows us to observe the tunneled electron without re-collision. As expected, we observe evidence of nodal planes from the orbitals of O$_2$ and C$_6$H$_6$ in the photo-electron spectrum. In addition, we find the nodal planes suppress ionization at certain alignment angles, allowing the neutral to survive longer. This leads to an enhancement in the high energy region of the photo-electron spectrum. In N$_2$ we also find a similar suppression and an enhanced high energy cutoff for molecules aligned perpendicular to the laser polarization vector.

\begin{figure}[t]
  \centering
  \includegraphics[width=\columnwidth]{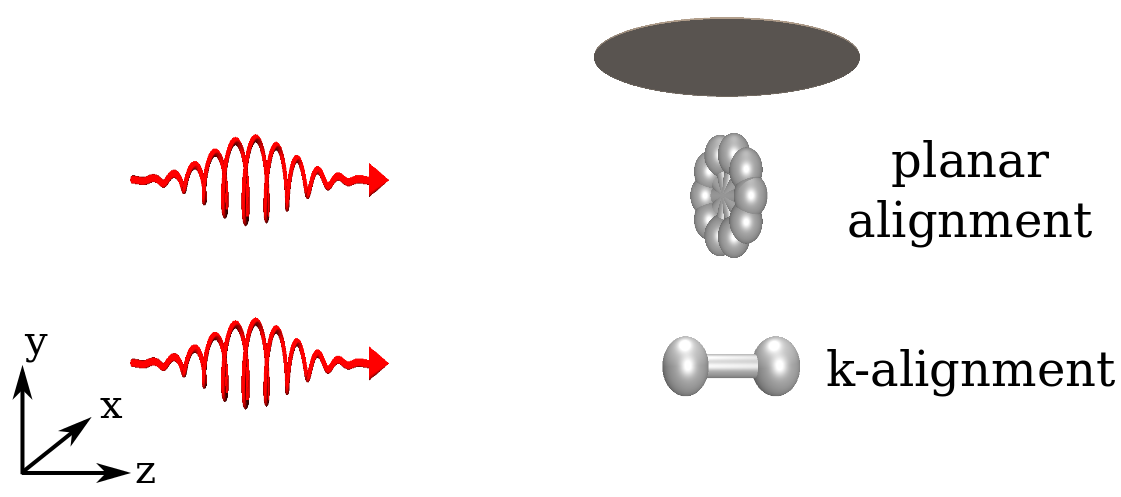}

  \caption{Molecular alignment conditions used in the experiment. The upper sketch shows a circularly polarized probe laser pulse focused onto molecules aligned in the plane of polarization. The lower sketch shows a probe pulse focused onto molecules aligned parallel to the propagation axis $z$ (called k-aligned). Experimentally the nature of the alignment distribution depends on the time delay between circularly polarized pump (not shown) and probe laser pulses. The 2D detector (top) records photo-electron spectra from single ionization under either planar or k-aligned conditions.}
  \label{fig:concept}
\end{figure}

\begin{figure*}[t]
  \centering

  \includegraphics[height=4.8cm]{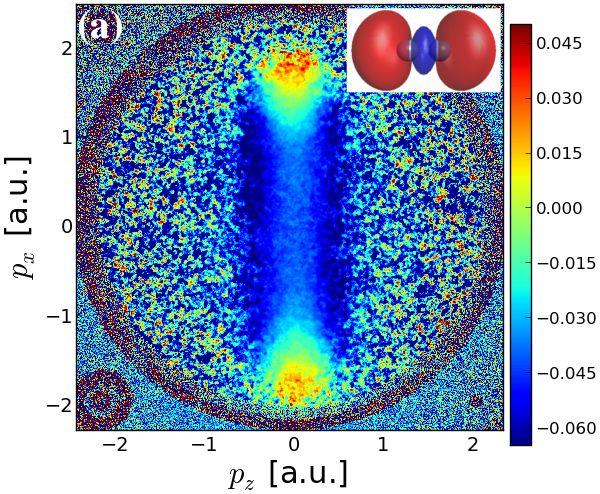}
  \includegraphics[height=4.8cm]{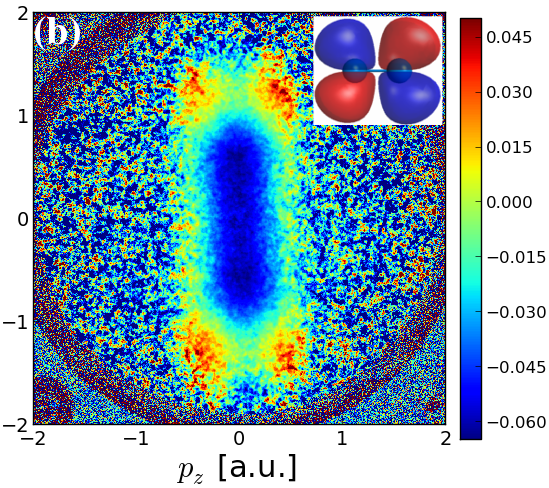}
  \includegraphics[height=4.8cm]{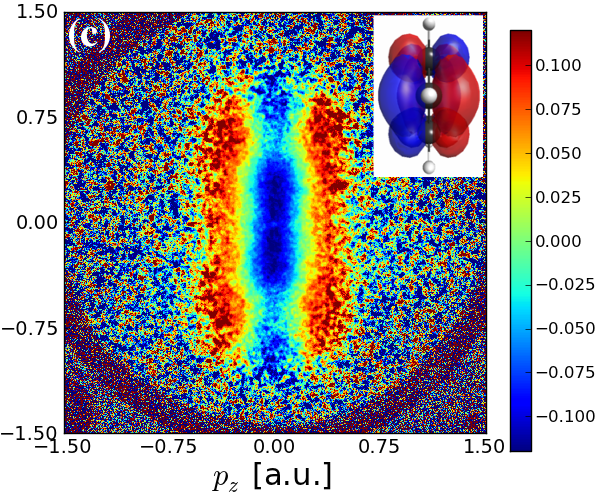}

  \parbox{0.32\linewidth}{\centering N$_2$}
  \parbox{0.32\linewidth}{\centering O$_2$}
  \parbox{0.32\linewidth}{\centering C$_6$H$_6$}
  \caption{The symmetry of the Dyson orbital (inset) enhances different regions of the measured normalized difference spectra. Normalized difference is $(A - B)/(A+B)$ where $A$ is the photo-electron spectrum measured at k-alignment (benzene: planar alignment) and $B$ at planar alignment (benzene: anti-alignment). (a) N$_2$ at $4.5 \times 10^{14}$ W/cm$^2$. (b) O$_2$ at $2.7 \times 10^{14}$ W/cm$^2$. (c) C$_6$H$_6$ at $4.0 \times 10^{13}$ W/cm$^2$.}

  \label{fig:expdiffs}
\end{figure*}

We use a circularly polarized aligning pulse from a Ti:Sapphire laser centred at 800 nm to align the molecules in the lab frame \cite{Smeenk2013}. The circularly polarized alignment (or pump) pulse produces a torque on the molecules, pulling the most polarizable axis (or plane in the case of C$_6$H$_6$) into the plane of polarization. This creates a rotational wavepacket that evolves under field-free conditions once the pump pulse has passed \cite{Stapelfeldt2003}. The wavepacket rephases every full revival period $T_R$, a value which depends solely on the rotational constants of the molecule. At $T_R$ the major axis (or plane) re-aligns in the plane of polarization: the $xy$ plane in Fig.~\ref{fig:concept}. We call this ``planar alignment'' \cite{Smeenk2013}.  At a different time delay during the wavepacket evolution, the molecules are anti-aligned. In the case of linear molecules, anti-alignment fixes the molecular frame to the propagation axis (k-axis) of the laser pulse, i.e.~the $z$ axis \cite{
Smeenk2013}. We call this ``k-alignment'' in Fig.~\ref{fig:concept}. The alignment dynamics of benzene, a planar molecule, can best be understood by concentrating on the normal vector to the plane of the benzene ring.  In a circularly polarized aligning pulse, the ring is initially pulled into the plane of polarization, that is, the normal vector aligns with the $z$-axis \cite{Smeenk2013}.  Anti-alignment of benzene corresponds to the normal vector de-localized in the plane of polarization ($xy$).

The circularly polarized pump pulse was stretched to $\sim 60$ fs for the diatomic molecules or 250 fs for benzene. The pump laser pulse is focused onto a pulsed, supersonic gas jet operating at 1 kHz. The jet temperature was estimated to be 5 K and we measured the degree of alignment $\langle \cos^2 \theta \rangle=0.7$ for N$_2$ and O$_2$. We control the nature of the molecular alignment by precisely controlling the time delay between alignment (or pump) and probe laser pulses \cite{Smeenk2013,Dooley03}. In N$_2$ we used the half revival (planar alignment: $4.07$ ps, k-alignment: $4.40$ ps). For O$_2$ we used the quarter revival for planar alignment ($2.90$ ps) and three-quarter revival for k-alignment ($8.68$ ps). Although we use different pump-probe timing values for N$_2$ and O$_2$, the molecular alignment distributions are the same for the two linear molecules. Lastly, for benzene we used the full revival for anti-alignment ($87.2$ ps) and planar alignment ($88.8$ ps). 

A 40 fs, circularly polarized probe pulse from the same laser system in the range $4\times10^{13} - 4\times10^{14}$ W/cm$^2$ singly ionizes the planar or k-aligned molecules. We will compare molecular frame photo-electron angular distributions from planar aligned molecules to k-aligned molecules.

We recorded photo-electron spectra using a velocity map imaging spectrometer (VMI) \cite{Eppink1997}. Because of the large inhomogeneous electric field in the spectrometer ($\approx 1 $ kV/cm) and the small mass of electrons, each measured spectrum is a two-dimensional projection of the 3D photo-electron spectrum. Each measurement contains $\sim 2 \times 10^7$ detected electrons. The spectra were corrected for dark noise by subtracting an image taken with the laser pulses blocked.

We observe alignment dependent ionization suppression for all three molecules. In N$_2$ there is a 7\% smaller signal from k-aligned molecules compared to planar aligned molecules. In O$_2$ the suppression is also 7\% while in benzene we found a 5\% smaller signal from planar aligned molecules than from anti-aligned molecules. A decreased ionization yield for one alignment vs.~another demonstrates that the geometry of the Dyson orbital, $\psi_D$ \cite{suppinfo,Patchkovskii2006} modifies the tunnel ionization rate \cite{Tong2002,MuthBohm2000}.

Qualitative features in the raw photo-electron spectra are similar for both planar and k-aligned molecules. To clearly observe the changes, we take a normalized difference, $ND = (A - B)/(A + B)$ where $A$ and $B$ are the k-aligned (benzene: planar aligned) and planar aligned (benzene: anti-aligned) distributions respectively. The normalized difference partially compensates for the large dynamic range in our measurement, allowing us to observe small modulations due to different orientations of the Dyson orbital. It also removes the influence of  inhomogeneous sensitivity of our microchannel plate detector.

The normalized difference spectra in Fig.~\ref{fig:expdiffs} each show two important features. Each normalized difference spectrum peaks at high $p_x$ momentum. This is due to angle-dependent saturation which is higher for alignment $A$ than for alignment $B$. The spectra also show suppression along nodal planes in the Dyson orbitals  of O$_2$ (Fig.~\ref{fig:expdiffs}b) and C$_6$H$_6$ (Fig.~\ref{fig:expdiffs}c). In Fig.~\ref{fig:expdiffs} the laser propagates along the horizontal axis ($z$) and the plane of polarization is $xy$. K-aligned molecules are therefore aligned along $z$ and planar aligned molecules lie in the $xy$ plane. Nodal planes in k-aligned O$_2$ and planar aligned C$_6$H$_6$ suppress ionization at low lateral momentum ($p_z \approx 0$). In all cases the ionization probability of the raw spectra drops by $10^{4}$ at larger lateral momenta  ($|p_z| > 0.8$). Consequently, the normalized difference signal is lost to experimental noise. The loss of signal in the lateral direction is 
characteristic of re-collision free tunneling \cite{Arissian2010}.

To place the enhancement of the photo-electron cutoff due to angle-dependent saturation on a more quantitative footing, we take the ratio of the k-aligned/planar-aligned spectra (benzene: planar-aligned/anti-aligned). A ratio of 1 corresponds to equal ionization probability from each alignment direction. A value greater than 1 represents enhancement of k-aligned (benzene: planar aligned) molecules. The distributions were first integrated from $-0.5 \leq p_z \leq 0.5$ before analyzing the ratio in the plane of polarization.

\begin{figure}[t]
  \centering
  \includegraphics[width=\columnwidth]{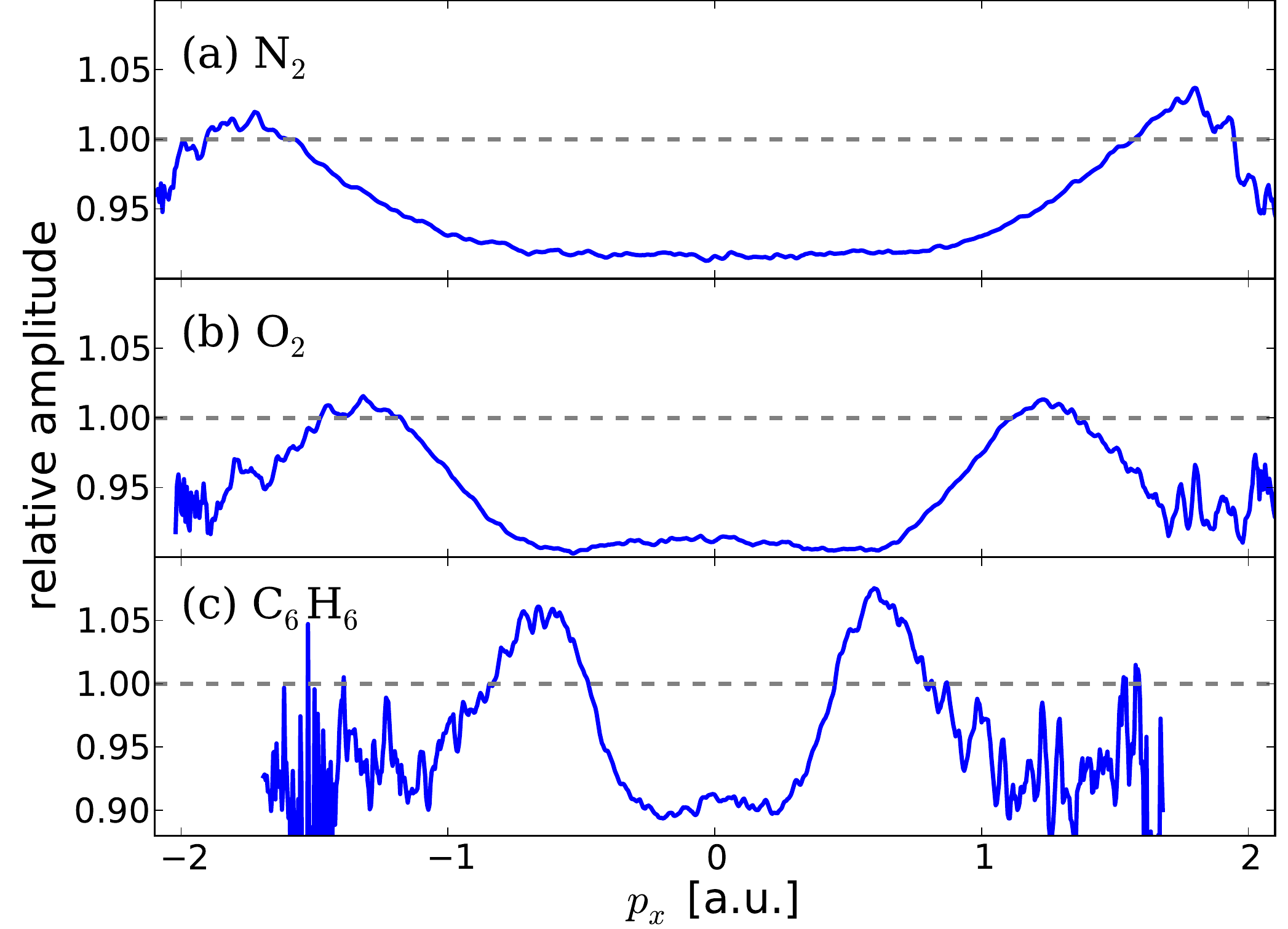}
  \caption{Enhancement of the photo-electron cutoff region due to angle dependent ionization. (a) k-aligned/planar-aligned spectra for N$_2$ at $4.5 \times 10^{14}$ W/cm$^2$ (b) k-aligned/planar-aligned spectra for O$_2$ at $2.7 \times 10^{14}$ W/cm$^2$ (c) planar-aligned/anti-aligned spectra for benzene at $4.0 \times 10^{13}$ W/cm$^2$}
  \label{fig:enhancementProfile}
\end{figure}

The ratio shows that the high momentum cutoff is enhanced for alignment $A$ for all three molecules (Fig.~\ref{fig:enhancementProfile}). This occurs whenever the angle dependent ionization rate is suppressed. In N$_2$ and O$_2$, k-aligned molecules have a lower ionization rate than planar aligned molecules \cite{Litvinyuk2003,Pavicic2007}. Figs.~\ref{fig:enhancementProfile}a and \ref{fig:enhancementProfile}b show the low momentum region is suppressed for k-aligned molecules while the high momentum region shows a slight enhancement. Beyond the maximum momentum the signal is lost to noise. In benzene, the spectrum shows suppression for planar aligned molecules in the low momentum region and enhancement in the cutoff (Fig.~\ref{fig:enhancementProfile}c). In the following we show that the high energy enhancement is a general feature and is captured by a simple tunneling model using the ground ion state and angle-dependent ionization rate.

\begin{figure*}
  \centering
  \includegraphics[width=0.32\linewidth]{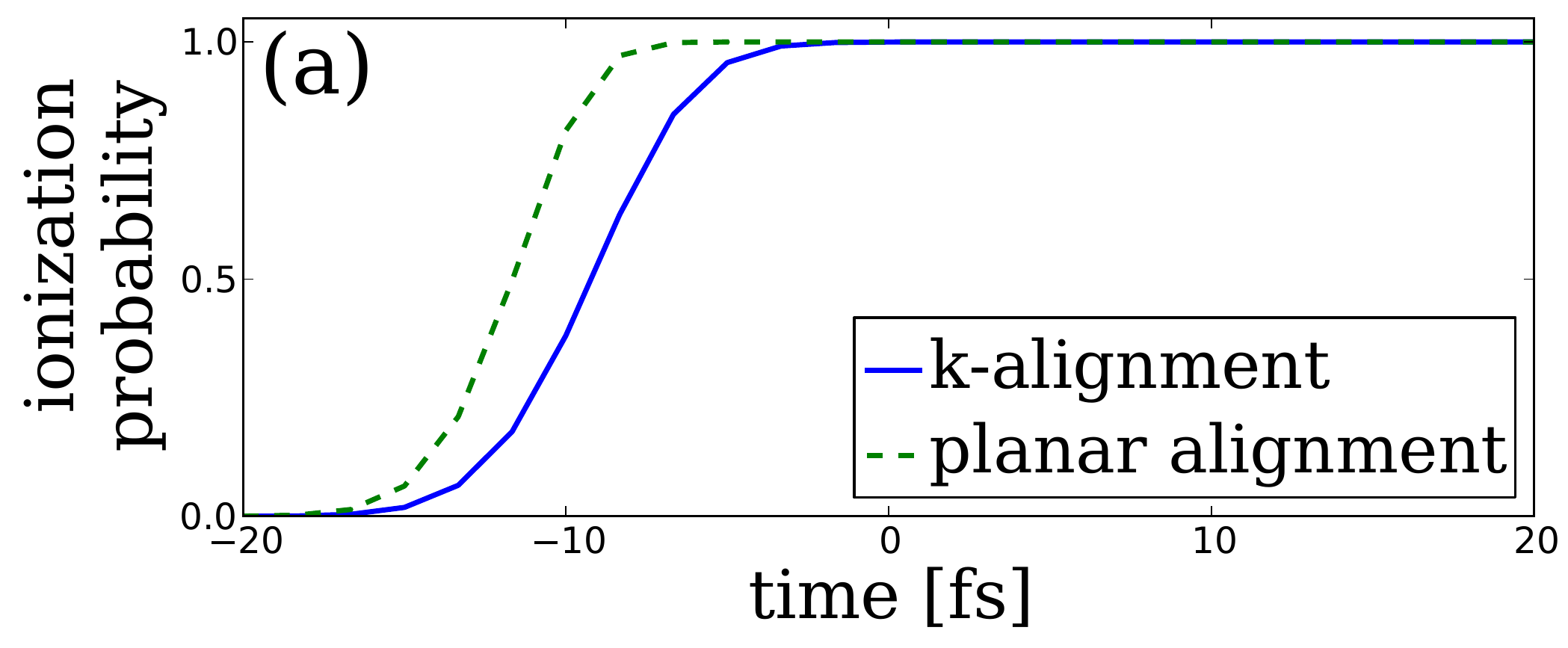}
  \includegraphics[width=0.32\linewidth]{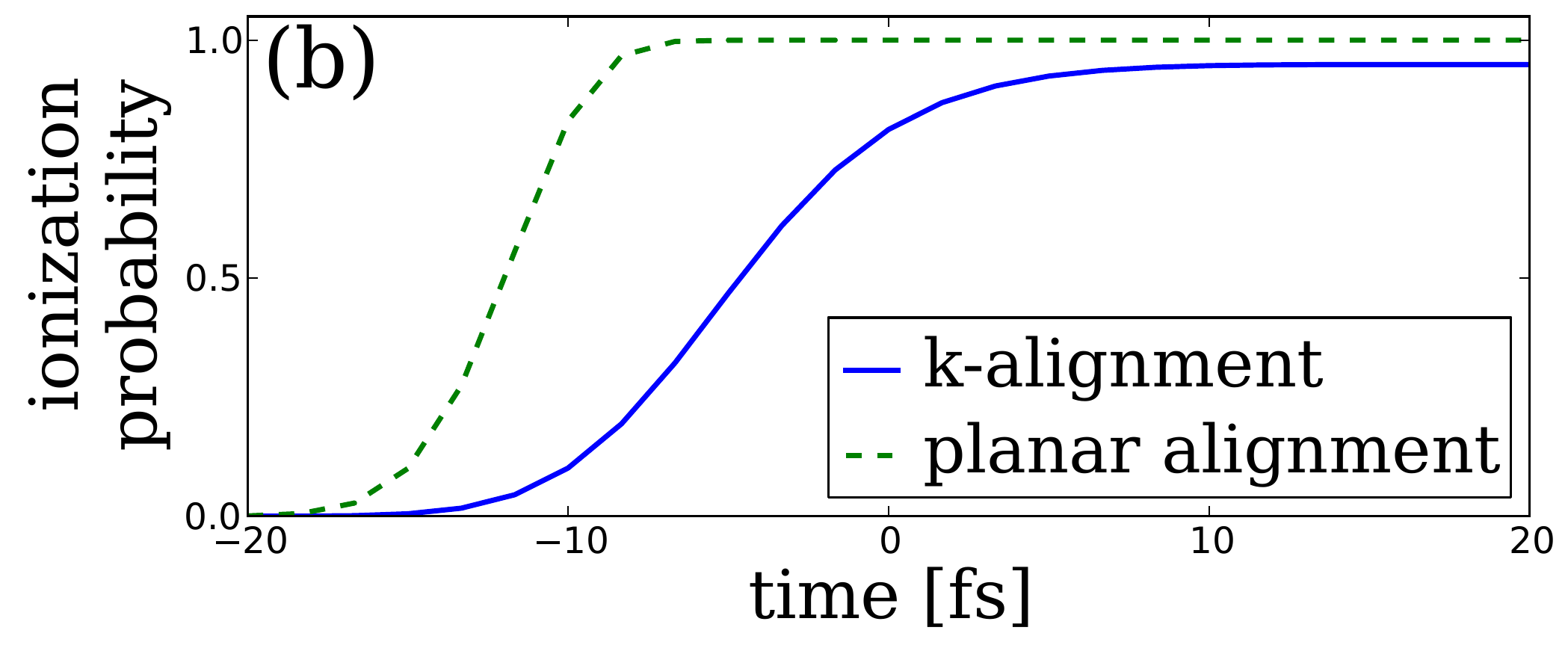}
  \includegraphics[width=0.32\linewidth]{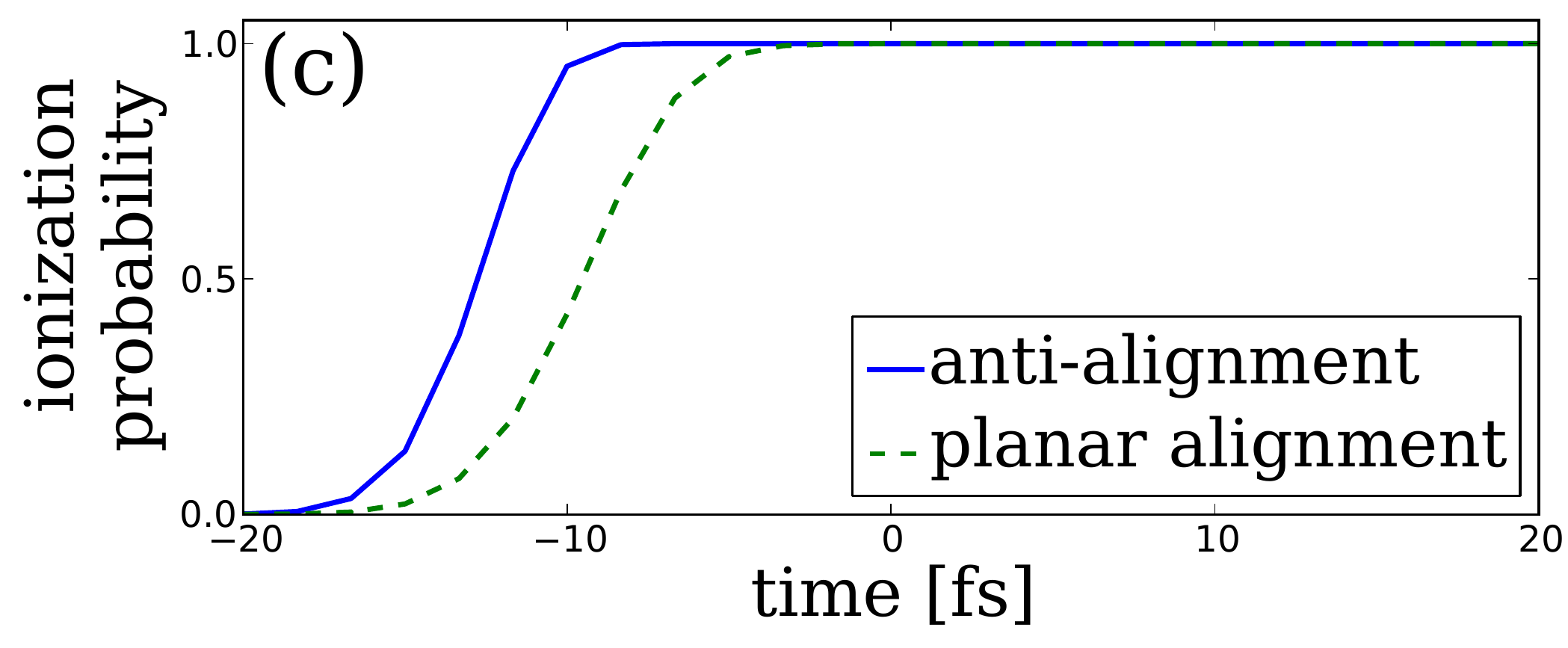}

  \includegraphics[width=0.32\linewidth]{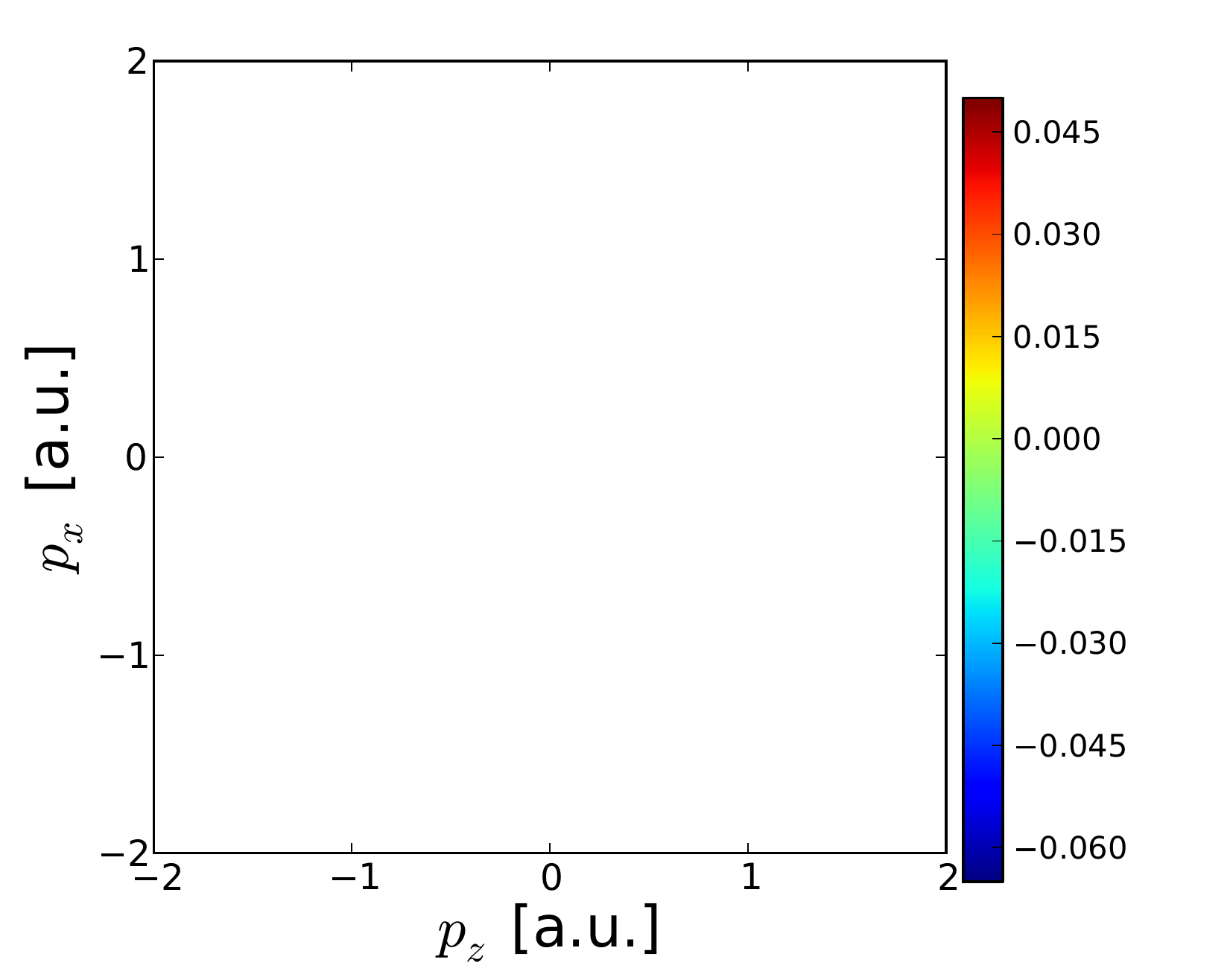}
  \includegraphics[width=0.32\linewidth]{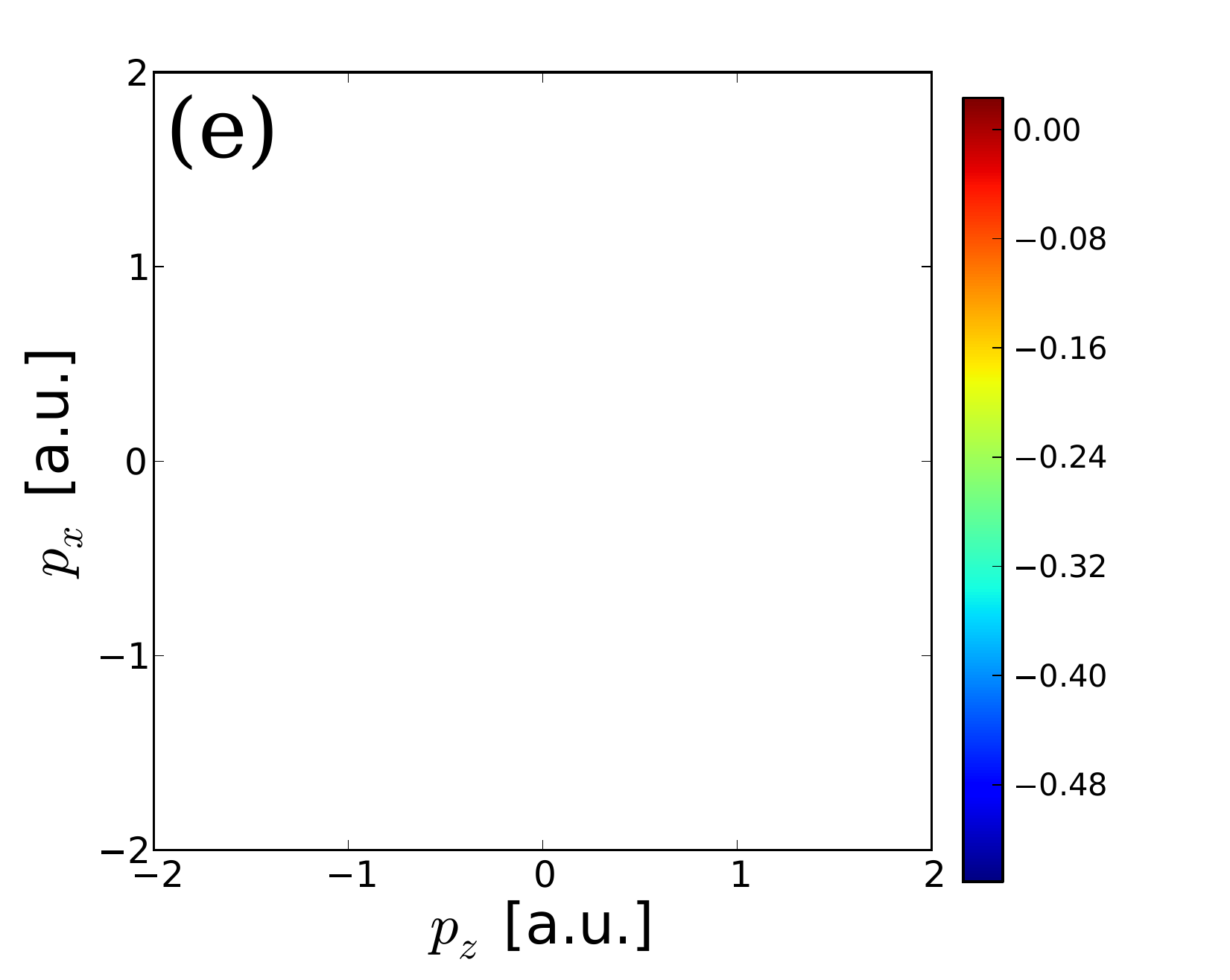}
  \includegraphics[width=0.32\linewidth]{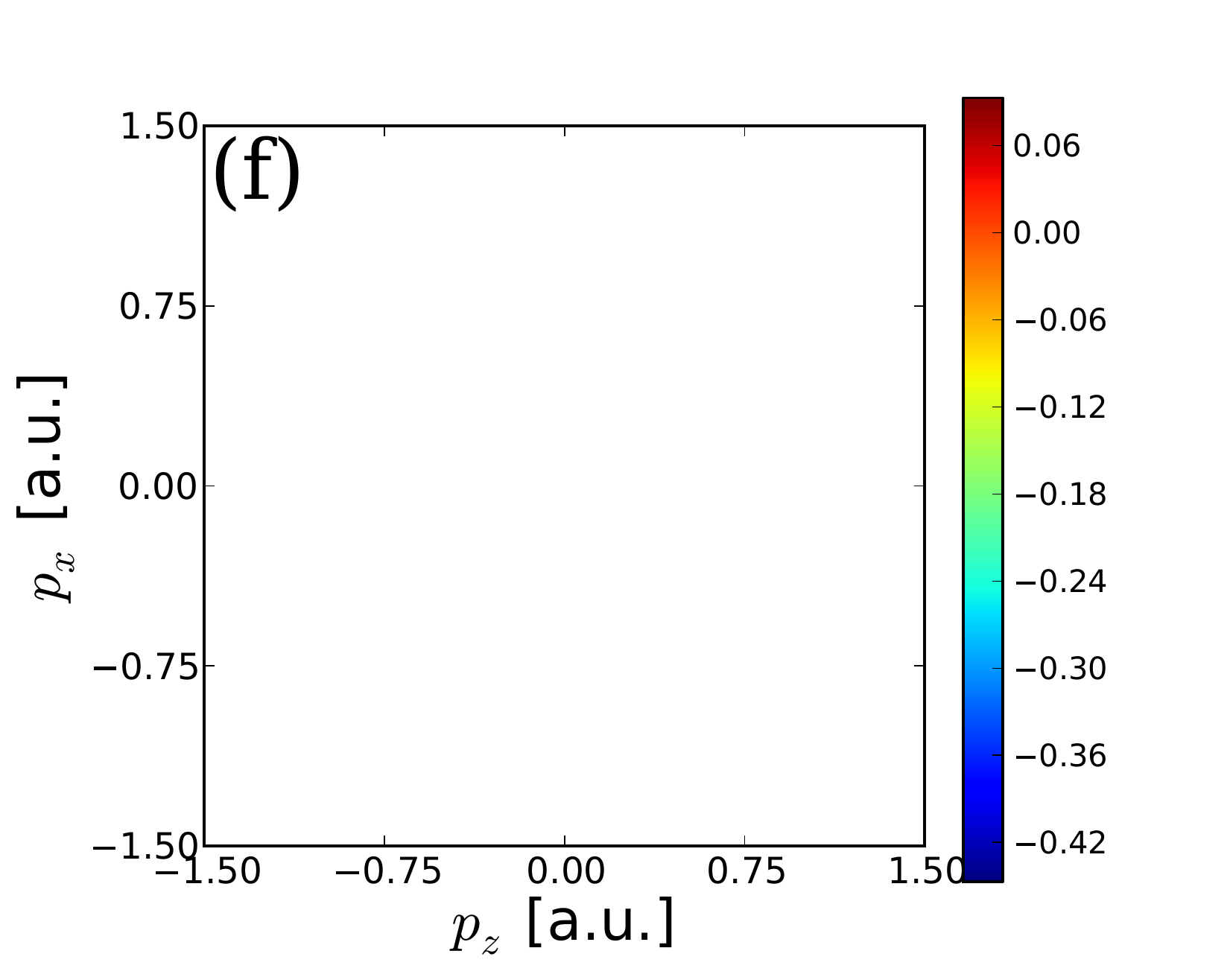}

  \parbox{0.32\linewidth}{\centering N$_2$}
  \parbox{0.32\linewidth}{\centering O$_2$}
  \parbox{0.32\linewidth}{\centering C$_6$H$_6$}
  \caption{{\bf Upper row:} Survival probability of the neutral depends on molecular alignment.  For (a) N$_2$ and (b) O$_2$,  k-aligned neutral molecules survive later in the pulse.  For (c) benzene, planar aligned neutral molecules survive later. {\bf Lower row:} Calculated normalized difference spectra illustrate how the angle dependent survival probability affects the photo-electron spectrum. (a),(d): N$_2$ at $4.5 \times 10^{14}$ W/cm$^2$.  (b),(e):  O$_2$ at $2.7 \times 10^{14}$ W/cm$^2$. (c),(f) C$_6$H$_6$ at $4.0 \times 10^{13}$ W/cm$^2$. 1\% Gaussian noise was added to each calculated spectrum prior to processing.}

  \label{fig:calcdiffs}
\end{figure*}

To begin, we calculate an ab-initio 3D momentum-space Dyson orbital $\psi_D ( \bsym{p})$.  To calculate the angle dependent tunneling rate, we use a modified version of a molecular tunneling theory. We write the instantaneous ionization rate as
\begin{multline}
  W\left( E(t), \beta(t) \right) = W_{ADK}\left(E(t), I_p(\beta(t)) \right) \times \alpha(\beta(t))  \times \\  \int \ud p_\perp \ud p_\parallel \left| \psi_D (\bsym{p}, \beta(t)) \right|^2 \exp \left(-p_\perp^2 \frac{\sqrt{2 I_p(\beta(t))}}{E(t)} \right)
  \label{eq:tunnelRateMod}
\end{multline}
where $\beta(t)$ is the angle between the laser polarization vector and the major polarizability axis in the molecule, $I_p(\beta(t))$ is the Stark shifted ionization energy \cite{suppinfo},  $W_{ADK} (E, I_p(\beta(t)))$ is the atomic tunnel ionization rate \cite{Chang2011}, $\alpha(\beta(t))$ is a free scalar parameter and the final integral factor includes the effect of the Dyson orbital, $\psi_D$. For our experiments using circularly polarized light, the angle $\beta$ changes at the optical frequency as the polarization vector sweeps around the molecular axes. Eq.~\eqref{eq:tunnelRateMod} is similar to many theories on tunnel ionization of molecules which split the rate into an atomic tunneling part and a geometrical part \cite{Tong2002,MuthBohm2000,Murray2011}.

The most prominent effect of including the Stark shift is to reduce the overall ionization rate (see online material \cite{suppinfo}). This is a result of the sensitive dependence of the tunnel ionization rate on the binding energy in the factor $W_{ADK}(E(t),I_p(\beta(t)))$. On the other hand, the angle-dependent Stark shift does not qualitatively alter the angular dependence \cite{suppinfo}. In the case of N$_2$, Eq.~\eqref{eq:tunnelRateMod} using $\alpha(\beta(t))=\textrm{const.}$ predicts the ionization rate changes very little with respect to angle (online material Fig.~1 \cite{suppinfo}). This is inconsistent with the well-known suppression of ionization perpendicular to the internuclear axis for N$_2$ \cite{Litvinyuk2003,Pavicic2007}. In our calculations we therefore set $\alpha(\pi/2) = \alpha(0)/4$ for N$_2$ only to correctly capture the ionization suppression for perpendicularly aligned N$_2$. This necessary modification to Eq.~\eqref{eq:tunnelRateMod} cannot be explained by ionization to 
excited ion states \cite{McFarland2008,Le2009} or, as we have demonstrated, by angle dependent Stark shifts.

The ionization probability from the Dyson orbital is given by $ P (t) = 1 - \exp \left[ - \int_{-\infty}^t W( E(t^\prime), I_p(\beta(t^\prime)) ) \ud t^\prime \right] $ and the instantaneous ionization yield is $\ud P/\ud t$. We use the instantaneous yield to weigh the contribution from a particular alignment angle as a function of time in the 40 fs pulse. At each time step we average the yield over all alignments present in the molecular distribution to calculate the net yield. The yield depends on the laser field $E$, the alignment of the bound orbital with respect to the polarization axis, and the remaining population in the Dyson orbital.

For N$_2$  and  O$_2$  the ionization rate from k-aligned molecules is lower than for planar aligned molecules (Figs.~\ref{fig:calcdiffs}a and \ref{fig:calcdiffs}b). This means that population survives in the neutral k-aligned molecules until later in the pulse when the laser field is larger. Consequently the photo-electron distribution will be shifted to slightly larger drift momentum $p=E(t_i)/\omega$ for k-aligned N$_2$ and O$_2$. This creates the enhancement in the high energy cutoff region shown in Figs.~\ref{fig:enhancementProfile}a and \ref{fig:enhancementProfile}b.  For benzene, the nodal plane along the molecular axis means that planar aligned molecules have a lower ionization rate and hence saturate later in the pulse (Fig.~\ref{fig:calcdiffs}c). Similar to k-aligned linear molecules, the longer survival of planar aligned benzene molecules enhances the high momentum region of the spectrum as we observed experimentally (Fig.~\ref{fig:enhancementProfile}c).

Figs.~\ref{fig:calcdiffs}d -- \ref{fig:calcdiffs}f show calculated two dimensional normalized difference spectra for the three molecules. These figures indicate an interplay between the saturation, which enhances the high momentum region, and the orbital geometry which restricts the photo-electron distribution in the lateral direction, $p_z$. In N$_2$ (Fig.~\ref{fig:calcdiffs}a), the neutral k-aligned molecules survive to higher intensity portions of the pulse. Since k-aligned N$_2$ has maximum electron density at $p_z = 0$, the photo-electron distribution is enhanced around $p_z=0$ and at large longitudinal momentum, $|p_x| \approx 1.8$ (cfr.~Fig.~\ref{fig:calcdiffs}d, Fig.~\ref{fig:expdiffs}a). The high energy enhancement is solely due to the alignment-dependent saturation intensity of the neutral. 

In O$_2$ (Fig.~\ref{fig:calcdiffs}b), the neutral k-aligned molecules also survive longer in the pulse, but the Dyson orbital has a nodal plane at $p_z = 0$. This shifts the enhancement in Fig.~\ref{fig:calcdiffs}e to non-zero lateral momentum, $|p_z| \approx 0.3$. In Figs.~\ref{fig:expdiffs}b and \ref{fig:calcdiffs}e we observe four bright red enhancement features around $p_z=0.3$, $p_x=1.3$ in the O$_2$ normalized difference distribution. 

Lastly, the benzene spectrum (Fig.~\ref{fig:calcdiffs}f) is similar to O$_2$. Since planar aligned benzene survives longer in the pulse (Fig.~\ref{fig:calcdiffs}c), it is a source of more high momentum electrons, as we saw in Fig.~\ref{fig:enhancementProfile}c. Similar to O$_2$, the maxima in the benzene spectrum are found at non-zero lateral momentum (Fig.~\ref{fig:calcdiffs}f). This is again a consequence of the nodal plane along $p_z=0$ for planar aligned benzene molecules that suppresses ionization in the plane of polarization.

The qualitative agreement between Figs.~\ref{fig:expdiffs} and \ref{fig:calcdiffs} is striking. Similar behaviour will be characteristic of most molecules. Since the tunnel ionization rate always depends on molecular alignment, molecules lying at difficult to ionize angles will always survive longer and produce the highest energy photo-electrons. While we have used circular polarization to concentrate on the photo-electron cut-off energy from strong field ionization, similar effects will appear in all re-scattering experiments. Cut-off harmonics that are produced by high momentum re-scattering electrons have been used to identify excited ion states and electron dynamics in high harmonic XUV spectra \cite{Smirnova2009a,McFarland2008,Le2009}. Our results suggest the laser intensity must be kept low and pulse duration short to avoid confusing the appearance of excited ion states with angle-dependent-depletion of the neutral.

We acknowledge financial support from NSERC, CIPI, JSPS and The Welch Foundation (Award No. A-1547).


\end{document}